\documentclass[9pt,twocolumn,twoside]{osajnl}
\usepackage{amsmath,mathtools}
\usepackage{mathrsfs}

\journal{josab} 

\setboolean{shortarticle}{true}

\title{A general approach to model counterpropagating continuous variable entangled states in a lossy CROW}

\author[1,2,*]{Hossein Seifoory}
\author[2]{Marc M. Dignam}

\affil[1]{Department of Physics, University of Toronto, 60 St. George Street, Toronto, Ontario M5S 1A7, Canada}
\affil[2]{Department of Physics, Engineering Physics and Astronomy, Queen's University, Kingston, Ontario K7L 3N6, Canada}

\affil[*]{hossein.seifoory@utoronto.ca}

\begin{abstract}
We present a general approach to model an integrated source of counterpropagating continuous-variable entangled states based on a coupled resonator optical waveguide that is pumped by a classical pulsed source incident from above the waveguide. This paper is an extension of our previous work~(Ref.~\cite{PhysRevA.100.033839}), where we analytically investigated the generation and propagation of continues-variable entangled states in this coupled-cavity system in the presence of intrinsic loss. However, in this work, we employ a numerical method to implement the Schmidt decomposition method rather than pursuing analytical methods. We show that not only this gives us a much higher degree of freedom in choosing the pumping parameters which were not possible to investigate analytically, but also it enables us to go beyond some of the approximations we had made to derive analytical expressions before.       
\end{abstract}

\setboolean{displaycopyright}{true}

\begin{document}

\maketitle

\section{Introduction}
Quantum entanglement is at the core of quantum information and has potential  applications in quantum computation~\cite{Ladd2010}, quantum teleportation~\cite{PhysRevLett.70.1895,Bouwmeester1997}, quantum cryptography~\cite{RevModPhys.74.145}, and quantum dense coding~\cite{PhysRevLett.69.2881}.
Entanglement can be established using discrete variables (DV), such as the polarization of a photon, or using continuous variables (CV), such as the quadrature components of beams of light. However, although DV systems provide high-fidelity operations, implementation of DV entanglement in photonics domain currently suffers from difficulties in single photon generation and detection, and from loss in integrated on-chip systems. CV entanglement, on the other hand, can be efficiently created and used for implementation of CV quantum protocols~\cite{RevModPhys.77.513,Huang2016,PhysRevA.83.042312,PhysRevX.5.041010,PhysRevLett.93.250503,PhysRevA.80.050303}. One of the advantages of using CV entanglement is that it is in general more robust to loss than its DV counterpart. \par
One of the processes which can be used to generate quantum correlated states, both in discrete and continuous variable domains, is spontaneous parametric down conversion~(SPDC)~\cite{PhysRevLett.75.4337,PhysRevLett.97.223602,PhysRevA.74.013815,Yang:07}. Briefly, in this method, which is a second order nonlinear process, a pump photon is converted into a signal and an idler photon~\cite{introductory}. SPDC has been implemented in both bulk media and integrated photonic structures. It has been shown that photon pairs can be generated via SPDC in integrated on-chip systems using materials with high second-order nonlinear susceptibility, such as AlGaAs~\cite{PhysRevA.85.013838} or AlN~\cite{Guo2017}. Different approaches have been followed to generate quantum correlated photons propagating in opposite directions in photonic platforms. Two examples are nonlinear periodic waveguides with horizontal pumping~\cite{PhysRevLett.118.183603} and ridge waveguides with vertical pumping~\cite{Orieux:11,PhysRevA.85.013838}. \par            

Unfortunately, most of the nonclassical attributes of light, such as squeezing and entanglement, are fragile with respect to loss~\cite{PhysRevLett.55.2409,Jasperse:11,Seifoory:17,PhysRevA.97.023840,PhysRevA.85.052330}, which can manifest as scattering, material absorption, or thermal noise. In optical systems, scattering loss is an unavoidable effect that arises from coupling to the environment. The destructive role of loss becomes more significant when working with integrated on-chip systems where surface roughness or limits on 3D confinement of the light can lead loss of photons from the system. Therefore, it is important to fully explore and understand the effect of loss on the generation and evolution of the nonclassical properties of light in integrated platforms. \par

In our previous work, we studied the generation and evolution of counterpropagating CV entangled states in coupled cavity structures~\cite{PhysRevA.97.023840,8819586,PhysRevA.100.033839}. We focused on photonic crystal~(PhC) based coupled-resonator optical waveguides~(CROWs), which consist of optical cavities weakly coupled in one dimension~\cite{Yariv:99,Ma:13,doi:10.1063/1.2737430,Takesue2014}. We considered vertical pumping, of a few cavities, which leads to counterpropagating entangled states, with no co-propagating pump at the outputs. One nice feature of CROWs is that by modifying the nature of the cavities and the separation between the cavities, one can adjust the dispersion and even the loss to some degree to optimize the system for a particular application~\cite{Yariv:99,Ma:13,doi:10.1063/1.2737430}. Employing a tight-binding~(TB) approximation, we were able to derive analytic time-dependent expressions for the number of photons in each cavity, as well as for the correlation variance between the photons in different pairs of cavities, to evaluate the degree of quantum entanglement. In addition, we were able to specify and model the effects of intrinsic scattering loss on the generated CV entanglement as a function of propagation distance, which is important for any application.\par
In deriving the analytical results in our previous work~\cite{8819586}, we limited ourselves to the cases in which (1) the central frequency of the pump was carefully chosen such that the group velocity dispersion of the CROW modes was close to zero in the vicinity of the generated signal and idler frequencies. And (2) the pump was considered to be Gaussian in time and frequency with a specific pumping configuration. These constraints allowed us to evaluate the biphoton wave function analytically. In this work, instead of deriving analytical expressions for the quantities of interest such as the number of photons in each cavity and the correlation variance between the photons in different pairs of cavities, we employ a numerical method to decompose the biphoton wave function, which allows us to overcome the two restrictions we had set, and present the numerical results. Moreover, not only such a numerical method allows us to go beyond the nearest-neighbor TB approximation, but also it allows us to study more general systems and pumping configurations.\par

This paper is organized as follows. In Section~\ref{subsec:generation} we first present the general theory of the generation the generalized two mode squeezed state in lossy CROWs using the backward Heisenberg method. We then consider the special case of a
pump that is Gaussian in time and space, and study the time dependence of the number of photons and the CV correlations. In Sections~\ref{subsec:Evolution} we present the we summarized the theory of the evolution of nonclassical states of light in a coupled-cavity system developed in our previous works. In Section~\ref{subsec:SVD} we use singular value decomposition to numerically study the number of photons in each cavity and the correlation variance between the photons in different pairs of cavities. In Section~\ref{sec:result} we first compare the numerical results with analytical results and then explore cases which were not possible to study analytically. Finally, in Section~\ref{sec:comclusion}, we present our conclusions.

\section{Theory}
\label{sec:theory}
\begin{figure}
\includegraphics[width=\linewidth]{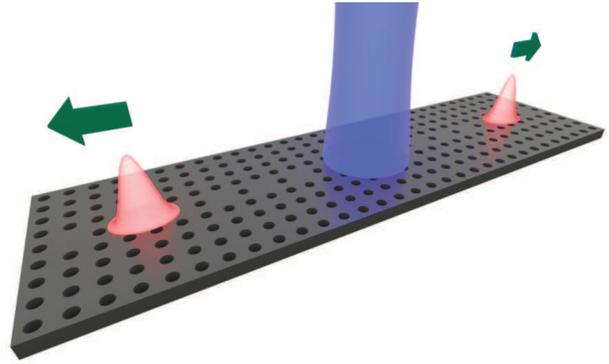}
\caption{Schematic picture of the particular CROW structure with period $D$ formed from defects in a slab PhC with a square lattice of period $d$ and height $h$. The blue region shows the region covered by the pump. The two red pulses indicate the generated signal and idler pulses. }
\label{fig:schematic}
\end{figure}

The system we consider here is a CROW structure with a pump pulse impinging normal to the plane on a set of central cavities from above. This system, which is schematically shown in Fig.~\ref{fig:schematic}, can be used to generate entangled signal and idler pulses. Because the pump is normal to the plane of the CROW, phase matching can only occur if the generated signal and idler modes propagate in opposite directions in the CROW structure as shown in Fig.~\ref{fig:schematic}. \par
Following the TB method~\cite{ashcroft2011solid,doi:10.1063/1.2737430}, which uses localized single-cavity modes as a basis, the generated light in such a coupled-cavity structure can be mathematically modeled as follows. Assuming that all the cavities are identical and support the same mode with complex frequency $\tilde{\omega}_F$, it has been shown~\cite{complex}  that in the nearest-neighbor tight-binding~(NNTB) approximation the dispersion relation can be written as
\begin{align}
\tilde{\omega}_{Fk}&\approx\tilde{\omega}_F[1-\tilde{\beta}_1 \cos(kD)]\nonumber\\
&\equiv\omega_{Fk}-i\gamma_{Fk},
\label{eq:dispersion_NNTB}
\end{align}
where $ \tilde{\beta}_1 $, $D$, and $k$ are respectively the complex coupling parameter, the periodicity of the CROW, and the Bloch vector component.  The imaginary part of the complex frequency ($\gamma_{Fk}$) is associated with the loss of the Bloch modes in the CROW. It is clear from Eq.~(\ref{eq:dispersion_NNTB}) that these modes experience different loss rates. We will discuss the implications of this when presenting our results.\par
In our calculation of the generation and evolution of entangled states in the system, we divide the analysis into two separate tasks. First, we employ the backward Heisenberg method~\cite{PhysRevA.77.033808}, which is intrinsically a lossless approach, to calculate the generation of entangled photons via SPDC method. Second, using the TB method and implementing Schmidt decomposition numerically, we calculate the evolution of the generated entangled states in the presence of loss. Note that ignoring the loss in the generation process and dividing the problem into two tasks is valid as long as the pump pulse is short enough in time that the loss in the generated modes is negligible over its duration. 
\subsection{Generation}
\label{subsec:generation}
In this section, we outline our implementation of the backward Heisenberg approach of Yang~\textit{et al.}~\cite{PhysRevA.77.033808}. The asymptotic out state for the \textit{generated} photons is given by 
\begin{equation}
\label{assym_out}
\left|\psi_{\text{out}}^{F}\right\rangle =e^{\frac{\beta}{\sqrt{2}}\int\text{d}k_{1}\text{d}k_{2}\,\phi\left(k_{1},k_{2}\right)\hat{b}_{k_{1}}^{\dagger}\hat{b}_{k_{2}}^{\dagger}-\text{H.c.}}\left|\text{vac}\right\rangle ,
\end{equation}
where $\left|\text{vac}\right\rangle$ is the vacuum state for the signal and idler modes, $\phi\left(k_{1},k_{2}\right)$ and $\hat{b}^\dagger_k$ are the biphoton wave function and the creation operators of the generated modes, respectively. The normalization constant $\beta$ is a real and positive quantity chosen to ensure that $\int \text{d}k_1\,\text{d}k_2\,|\phi\left(k_{1},k_{2}\right)|^{2}=1$. Considering that we are interested in counterpropagating entangled states, we will always choose the pump parameters such that to a very good approximation, $\phi(k_1,k_2)$ is nonzero only when $k_1$ and $k_2$ have opposite signs The integrals run over the first Brillouin zone $-\pi/D \leq k \leq \pi/D$ of the CROW.\par

As we shall see, due to the requirements of phase matching in the plane of the CROW, unless the pump pulse is very short in time and only excites one or two cavities, the generated photon pairs will have $k$-vectors with opposite signs. To encode this, we write the biphoton wave function as
\begin{equation}
\label{eq:Luke_Phi}
\Phi\left(k_{1},k_{2}\right)\equiv\sqrt{2}\phi\left(k_{1},k_{2}\right)\Theta\left(k_{1}\right)\Theta\left(-k_{2}\right),
\end{equation}
where $ \Theta\left(k\right) $ is the Heaviside function.\par
 Employing a Schmidt decomposition~\cite{peres2006quantum,doi:10.1119/1.17904}, we have
 \begin{equation}\label{eq:phi_inermsof_mu_nu}
\Phi\left(k_{1},k_{2}\right)=\sum_{\lambda}\sqrt{p_{\lambda}}\mu_{\lambda}\left(k_{1}\right)\nu_{\lambda}\left(k_{2}\right),
 \end{equation}
for $p_{\lambda} > 0$ with $\sum_{\lambda}p_{\lambda}=1$, where the Schmidt functions are orthonormal, 
\begin{equation}
\int\text{d}k\mu_{\lambda}(k)\mu^*_{\lambda^\prime}(k)=\int\text{d}k\nu_{\lambda}(k)\nu^*_{\lambda^\prime}(k)=\delta_{\lambda,\lambda^{\prime}}.
\end{equation}
We extend the sets of ${\mu_\lambda(k)}$ and ${\nu_\lambda(k)}$ associated with $p_\lambda>0$ to form complete sets with 
\begin{equation}
    \label{eq:orthonormal}
    \sum_\lambda \mu_\lambda(k)\mu^*_\lambda(k^\prime)=\sum_\lambda \nu_\lambda(k)\nu^*_\lambda(k^\prime)=\delta(k-k^\prime),
\end{equation}
with some of the $p_\lambda$ appearing in Eq.~(\ref{eq:phi_inermsof_mu_nu}) then equal to zero. The generated squeezed state can be written as
\begin{equation}
\left|\psi_{\text{out}}^{F}\right\rangle =\hat{S}\left|\text{vac}\right\rangle ,
\end{equation}  
where, from Eqs.~(\ref{assym_out}), (\ref{eq:Luke_Phi}), and (\ref{eq:phi_inermsof_mu_nu}), the squeezing operator, $ \hat{S} $, is given by
\begin{align}
\label{eq:S_operator}
\hat{S}&=\exp\left(\beta\int\text{d}k_{1}\text{d}k_{2}\,\sum_{\lambda}\sqrt{p_{\lambda}}\mu_{\lambda}\left(k_{1}\right)\nu_{\lambda}\left(k_{2}\right)\right.\left.\hat{b}_{k_{1}}^{\dagger}\hat{b}_{k_{2}}^{\dagger}-\text{H.c.}\vphantom{\beta\int\text{d}k_{1}\text{d}k_{2}\,\sum_{\lambda}}\right)\nonumber\\
&=\exp\left(\sum_{\lambda}r_{\lambda}\hat{B}_{\lambda}^{\dagger}\hat{C}_{\lambda}^{\dagger}-\sum_{\lambda}r_{\lambda}^{*}\hat{B}_{\lambda}\hat{C}_{\lambda}\right),
\end{align}
where $r_{\lambda}=\beta\sqrt{p_{\lambda}}$ is the squeezing parameter,
\begin{align}
\label{eq:B_operator}
\hat{B}_{\lambda}&\equiv\int\mu_{\lambda}^{*}\left(k\right)\hat{b}_{k}\text{d}k,
\end{align}
and
\begin{align}
\label{eq:C_operator}
\hat{C}_{\lambda}&\equiv\int\nu_{\lambda}^{*}\left(k\right)\hat{b}_{k}\text{d}k.
\end{align}
Using Eq.~(\ref{eq:orthonormal}), it can be shown that $ [\hat{B}_{\lambda},\hat{B}_{\lambda^{\prime}}^{\dagger}]=[\hat{C}_{\lambda},\hat{C}_{\lambda^{\prime}}^{\dagger}]=\delta_{\lambda,\lambda^{\prime}} $ and $ [\hat{B}_{\lambda},\hat{C}_{\lambda^{\prime}}^{\dagger}]=[\hat{B}_{\lambda},\hat{B}_{\lambda^{\prime}}]=[\hat{C}_{\lambda},\hat{C}_{\lambda^{\prime}}]=[\hat{B}_{\lambda},\hat{C}_{\lambda^{\prime}}]=0$. Note that \eqref{eq:S_operator} is simply a product of two-mode squeezing operators, where the modes are not the Bloch modes but rather are the Schmidt modes formed from combinations of Bloch modes.\par
The theory described thus far is independent of the temporal and spatial form of the pump pulse, as long as $\phi(k_1,k_2)$ is only nonzero when $k_1$ and $k_2$ have opposite signs. However, we now consider a practical pumping configuration in which the pump is Gaussian in time and space with $W_T$, $W_S$, and $\omega_S$ to be its frequency width related parameter, spot size at $x=0$, and central frequency, respectively. Here, rather than repeating the same calculations as we did in our previous work~\cite{PhysRevA.100.033839} to approximately obtain the biphoton wave function, we refer the reader to our previous work, where we derive the following expression for the biphoton wave function for this pump pulse:
\begin{equation}
\label{biphoton_02}
\begin{multlined}
{\Phi}(k_{1},k_{2})=Q_{0}\exp\left(\frac{-(k_{1}+k_{2})^{2}W_S^{2}}{4}\right)\\
\times\exp\left(-\left(\frac{2\omega_{F}-\beta_{1}\omega_{F}\left[\cos\left(k_{1}D\right)+\cos\left(k_{2}D\right)\right]-\omega_{P}}{2c}\right)^{2}W_T^{2}\right)\\
\times\Theta\left(k_{1}\right)\Theta\left(-k_{2}\right),
\end{multlined}
\end{equation}
where
\begin{equation}\label{eq:Q_0}
Q_{0}\equiv\frac{i\alpha\bar{\chi}_{2}}{\beta c}\sqrt{\frac{\hbar\omega_{F}^{2}\omega_{S}W_T}{\varepsilon_{0}\left(2\pi\right)^{3/2}}},
\end{equation}
$\omega_F$ is the generated signal and idler frequency, and $\bar{\chi}_{2}$ is the effective second-order susceptibility for the system. Because we shall always choose a pump frequency such that $\omega_P/2$ lies within the band of the CROW, for some wavevector, $k_o$, we can write the pump frequency in the form
\begin{equation}
\omega_{P}=2\omega_{Fk_{0}}=2\omega_{F}-2\beta_{1}\omega_{F}\cos\left(k_{0}D\right),
\end{equation}
where $\beta_1$ is the real part of $ \tilde\beta_1$. Using this in Eq.~(\ref{biphoton_02}), we obtain 
\begin{equation}
\label{biphoton_03}
\begin{multlined}
{\Phi}(k_{1},k_{2})=Q_{0}\exp\left(\frac{-(k_{1}+k_{2})^{2}}{2\mathscr{E}_+^2}\right)\\
\times\exp\left(-\frac{\left(\cos\left(k_{1}D\right)+\cos\left(k_{2}D\right)-2\cos\left(k_{0}D\right)\right)^2}{2\mathscr{E}_-^2}\right)\\
\times\Theta\left(k_{1}\right)\Theta\left(-k_{2}\right),
\end{multlined}
\end{equation}
where 
\begin{equation}
    \mathscr{E}_+\equiv\frac{\sqrt{2}}{W_S}
\end{equation}
and
\begin{equation}
    \mathscr{E}_-\equiv\frac{\sqrt{2}c}{\beta_1 \omega_F W_T}.
\end{equation}
\par

\subsection{Evolution}
\label{subsec:Evolution}
Following the formalism developed in our previous work on the evolution of nonclassical states of light in a coupled-cavity system~\cite{PhysRevA.97.023840}, the individual single-mode cavity annihilation operator , $ \hat{a}_p $, for the $ p^{th} $ cavity mode can be written as
\begin{equation}
\label{a_to_b}
\hat{a}_{p}(t)=\sqrt{\frac{D}{2\pi}}\int\hat{b}_{k}(t)e^{ikpD}\text{d}k,
\end{equation}
where $ \hat{b}_k $ is the $ k^{th} $ mode annihilation operator of the coupled-cavity-system. Solving the adjoint master equation~\cite{open_quantum_system}, we have previously shown that the time evolution of the full coupled-cavity annihilation operator after the pump pulse is gone is given by 
\begin{equation}
\label{b_time}
\hat{b}_k(t)=\hat{b}_ke^{-i\tilde{\omega}_{Fk}t},
\end{equation}
where $ \hat{b}_k=\hat{b}_k(0) $ is the corresponding operator in the Schr\"{o}dinger representation~\cite{PhysRevA.85.013809}. Using Eqs.~(\ref{a_to_b}), (\ref{b_time}), and their complex conjugates, the time dependent average photon number in the $ p^{th} $ cavity can be written as 
\begin{align}
\label{eq:apdap}
\left\langle \hat{a}_{p}^{\dagger}\left(t\right)\hat{a}_{p}\left(t\right)\right\rangle &=\frac{D}{2\pi}\int\int \text{d}k\text{d}k^{\prime}\left\langle \hat{b}_{k}^{\dagger}\hat{b}_{k^{\prime}}\right\rangle e^{-i\left(k-k^{\prime}\right)pD}\nonumber\\
&\times\left(e^{i\tilde{\omega}_{F}^{*}\left(1-\tilde{\beta}_{1}^{*}\cos(kD)\right)t}e^{-i\tilde{\omega}_{F}\left(1-\tilde{\beta}_{1}\cos(k^{\prime}D)\right)t}\right),
\end{align} 
where we have used the lossy dispersion relation of the CROW structure~[Eq.~(\ref{eq:dispersion_NNTB})]. 
 To facilitate the evaluation of $\left\langle \hat{b}_{k}^{\dagger}\hat{b}_{k^{\prime}}\right\rangle$, we introduce the restricted operators,
\begin{align}
\hat{b}_{k,+}&\equiv\Theta\left(k\right)\hat{b}_{k}\nonumber\\
\hat{b}_{k,-}&\equiv\Theta\left(-k\right)\hat{b}_{k}.
\end{align}
Using these operators, we can write 
\begin{equation}
    \label{eq:4_terms}
    \left\langle \hat{b}_{k}^{\dagger}\hat{b}_{k^{\prime}}\right\rangle=
    \left\langle \hat{b}_{k,+}^{\dagger}\hat{b}_{k^{\prime},-}^{}+
    \hat{b}_{k,-}^{\dagger}\hat{b}_{k^{\prime},-}^{}+
    \hat{b}_{k,+}^{\dagger}\hat{b}_{k^{\prime},+}^{}+
    \hat{b}_{k,-}^{\dagger}\hat{b}_{k^{\prime},+}^{}\right\rangle.
\end{equation}
To evaluate each of these terms, we use the following Bogoliubov transformations
\begin{equation}
\begin{multlined}
\label{bogol1}
\hat{S}^{\dagger}\hat{b}_{k,+}\hat{S}=\hat{S}^{\dagger}\sum_{\lambda}\mu_{\lambda}\left(k\right)\hat{B}_{\lambda}\hat{S}\\=\sum_{\lambda}\mu_{\lambda}\left(k\right)\left[\hat{B}_{\lambda}\cosh\left(r_\lambda\right)-\hat{C}_{\lambda}^{\dagger}\sinh\left(r_\lambda\right)\right], 
\end{multlined}
\end{equation}
\begin{equation}
\begin{multlined}
\label{bogol2}
\hat{S}^{\dagger}\hat{b}_{k,-}\hat{S}=\hat{S}^{\dagger}\sum_{\lambda}\nu_{\lambda}\left(k\right)\hat{C}_{\lambda}\hat{S}\\=\sum_{\lambda}\nu_{\lambda}\left(k\right)\left[\hat{C}_{\lambda}\cosh\left(r_\lambda\right)-\hat{B}_{\lambda}^{\dagger}\sinh\left(r_\lambda\right)\right].
\end{multlined}
\end{equation}
Using these in \eqref{eq:4_terms}, we obtain
\begin{equation}
\label{eq:bdkbk}
\left\langle \hat{b}_{k}^{\dagger}\hat{b}_{k^{\prime}}\right\rangle =\sum_{\lambda}\bigg(\mu_{\lambda}^{*}\left(k\right)\mu_{\lambda}\left(k^{\prime}\right)+\nu_{\lambda}^{*}\left(k\right)\nu_{\lambda}\left(k^{\prime}\right)\bigg)
\sinh^{2}\left(r_\lambda\right).
\end{equation} 
To study the degree of entanglement between the photons in cavities $ p $ and $ p^\prime $ in a CROW, we employ the sufficient inseparability criterion of Duan~\textit{et al.}, which is based on the sum of the variances and is defined as~\cite{PhysRevLett.84.2722,PhysRevLett.84.2726,Masada2015,Zhang2015}
\begin{equation}
\Delta_{p,p^\prime}^2=\left\langle [\Delta(\hat{X}_p-\hat{X}_{p^\prime})]^2\right\rangle+ \left\langle [\Delta(\hat{Y}_p+\hat{Y}_{p^\prime})]^2\right\rangle < 4,
\label{eq:correlation_nm_def}
\end{equation}  
where 
\begin{equation}
\label{X_and_Y}
\begin{aligned}
\hat{X}_p&\equiv\hat{a}_p+\hat{a}_p^{\dagger},\\
\hat{Y}_p&\equiv-i(\hat{a}_p-\hat{a}_p^{\dagger}). 
\end{aligned}
\end{equation}
Note that although in this work we only use the CV inseparability criterion of Duan~\textit{et al.}, there are some other sufficient inseparability criteria one can use to investigate CV entanglement~\cite{PhysRevLett.88.120401,PhysRevA.67.052104,PhysRevLett.96.050503} such as the criterion of Mancini~\textit{et al.}, which is based on the product of the same variances~\cite{PhysRevLett.88.120401}.\par
Using Eq.~(\ref{X_and_Y}) in Eq.~(\ref{eq:correlation_nm_def}), the time-dependent correlation variance can be written as 
\begin{equation}
\label{eq:entanglement_01}
\Delta_{pp^\prime}^{2}=4+4\left(\langle \hat{a}_{p}^{\dagger}\hat{a}_{p}\rangle +\langle \hat{a}_{p^{\prime}}^{\dagger}\hat{a}_{p^{\prime}}\rangle -\langle \hat{a}_{p}\hat{a}_{p^{\prime}}\rangle-\langle \hat{a}_{p}^{\dagger}\hat{a}_{p^{\prime}}^{\dagger}\rangle \right).
\end{equation}\par
Following a procedure similar to that used to arrive at Eqs.~(\ref{eq:apdap}) and (\ref{eq:bdkbk}), we derive the other expectation values that are needed to evaluate the variances of the quadrature operators and the correlation variance in the CROW structure~(see Ref.~\cite{PhysRevA.100.033839}).
\subsection{Singular Value Decomposition}
\label{subsec:SVD}
Here, rather than deriving analytical expressions for the quantities of interest such as the number of photons in each cavity and the correlation variance between the photons in different pairs of cavities, we employ a numerical method to decompose the biphoton wave function (\eqref{biphoton_03}) and present the numerical results. \par
Our aim is to determine the Schmidt decomposition of Eq.~\ref{eq:phi_inermsof_mu_nu} numerically. The first step to numerically implement the Schmidt decomposition method (Eq.~(\ref{eq:phi_inermsof_mu_nu})) is to discretize the biphoton wave function $\Phi(k_1,k_2)$. Therefore, we first put $k_1$ and $k_2$ on a grid such that $\Phi(k_1,k_2)$ becomes a $m\times n$ matrix, where $m$ and $n$ are determined by sizes of vectors $\overrightarrow{k}_1$ and $\overrightarrow{k}_2$. Let us denote this matrix by A for the moment. As shown in \eqref{eq:SVD}, using singular value decomposition~(SVD) method, it is possible to decompose a matrix $\overleftrightarrow{A}$ into the product of three matrices $\overleftrightarrow{A}=\overleftrightarrow{U}\overleftrightarrow{D}\overleftrightarrow{V}^T$, where $\overleftrightarrow{U}$ and $\overleftrightarrow{V}$ are orthonormal and $\overleftrightarrow{D}$ is a diagonal matrix with positive entries
\begin{widetext}
\begin{equation}
    \stackrel{\mbox{$\overleftrightarrow{A}$}}{ \begin{pmatrix}
  a_{1,1} & a_{1,2} & \cdots & a_{1,n} \\
  a_{2,1} & a_{2,2} & \cdots & a_{2,n} \\
  \vdots  & \vdots  & \ddots & \vdots  \\
  a_{m,1} & a_{m,2} & \cdots & a_{m,n} 
 \end{pmatrix}
 }
 \approx
  \stackrel{\mbox{$\overleftrightarrow{U}$}}{ \begin{pmatrix}
  u_{1,1} & \cdots & u_{1,r} \\
  \vdots  & \ddots &      \\
  u_{m,1} &      & u_{m,r} 
 \end{pmatrix}
 }
    \stackrel{\mbox{$\overleftrightarrow{D}$}}{ \begin{pmatrix}
  d_{1,1} & 0 &\cdots  \\
  0  & \ddots &      \\
  \vdots &      & d_{r,r} 
 \end{pmatrix}
 }
   \stackrel{\mbox{$\overleftrightarrow{V}^T$}}{ \begin{pmatrix}
  v_{1,1} & \cdots & v_{1,n} \\
  \vdots  & \ddots &      \\
  v_{r,1} &      & v_{r,n} 
 \end{pmatrix}
 }.
\label{eq:SVD}
\end{equation}
\end{widetext}In general, the SVD calculation can be thought of finding the eigenvectors and eigenvalues of $\overleftrightarrow{A}\overleftrightarrow{A}^T$ and $\overleftrightarrow{A}^T\overleftrightarrow{A}$. The two matrices $\overleftrightarrow{U}$ and $\overleftrightarrow{V}$, known as the right and left singular matrices and their columns are made up of the eigenvectors ("singular vectors") of $\overleftrightarrow{A}\overleftrightarrow{A}^T$ and $\overleftrightarrow{A}^T\overleftrightarrow{A}$, respectively. Thus we see that a matrix $\overleftrightarrow{A}$ can be decomposed in terms a sum of rank one matrices (vectors) as 
\begin{equation}\label{eq:SVD_A}
    \overleftrightarrow{A}=\sum^{min\{m,n\}}_id_i \, \overrightarrow{u}_i \, \overrightarrow{v}^T_i,
\end{equation}
where $d_i$ is the $i$th singular value and $\overrightarrow{u}_i$, $\overrightarrow{v}_i$ are the corresponding left and right singular vectors. Comparing Eqs.~(\ref{eq:phi_inermsof_mu_nu}) and (\ref{eq:SVD_A}), it can be seen that what we are generally referring to as the singular values, $d_i$, are actually equivalent to the square root of the Schmidt coefficients $p_\lambda$ and the $\overrightarrow{u}_i$ and $\overrightarrow{v}_i$ are the Schmidt modes on our $k$-grid.  \par
As we shall see shortly, applying the Schmidt decomposition numerically enables us to easily evaluate the evolution of the generated state in the presence of loss. \par

\section{Results}
\label{sec:result}
The CROW considered in this work consists of a dielectric slab of refractive index $n=3.4$ having a square array of cylindrical air voids of radius $a=0.4d$, height $h=0.8d$, and lattice vectors $\textbf{a}_1=d\hat{\textbf{x}}$ and $\textbf{a}_2=d\hat{\textbf{y}}$, where $d$ is the period. The cavities are point defects formed by periodically removing air voids in a line with $D=2d$ (see Fig.~\ref{fig:schematic}). The remaining physical parameters for this CROW are given in Ref.~\cite{doi:10.1063/1.2737430}. We use finite difference time domain (FDTD) calculations to calculate the mode fields and complex frequencies for the single-defect modes and then use these to calculate the TB parameters. The complex frequency, $\tilde{\omega}_F$, and the complex coupling parameter, $\tilde{\beta}_1$, of the structure are found to be $(0.305-i7.71\times10^{-6})4\pi c/D$, and $9.87\times10^{-3}-i1.97\times10^{-5}$, respectively.\par
\begin{figure}[htbp]
\centering
\includegraphics[width=\linewidth]{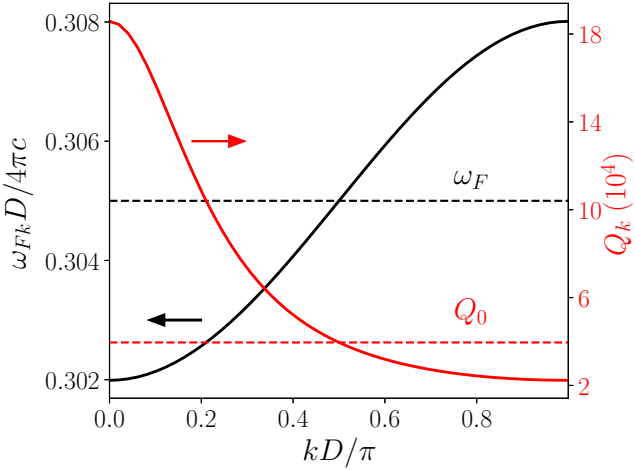}
\caption{Mode frequency (left axis) and quality factor (right axis) as a function of the Bloch vector for the CROW structure. The dashed black and red lines represent the resonant frequency and the quality factor for the individual cavity, respectively.}
\label{fig:dispersion}
\end{figure}

In this work, we ignore the extrinsic losses due to the fabrication imperfections, such as surface roughness on the circumference of an air void, and only consider the intrinsic losses originated from the out-of-plane radiation from the cavities. Note that the quality factor ($Q$) of CROWs is not solely determined by the $Q_k$ of the individual microcavities~($Q_k=\omega_{Fk}/2\gamma_k$). In fact, the interference of the out-of-plane radiation from the cavities affects the $Q$ of CROW consequently causing it to depend on the Boch vector. In Fig.~\ref{fig:dispersion}, we plot the mode frequency and quality factor as a function of the Block vector for our CROW structure. As can be seen, the group velocity and quality factor of the modes differ greatly across the first Brillouin zone. This wide variation in the loss of the different modes has a significant effect on the loss dynamics in the system, as we shall see.\par
\begin{figure}[!htb]
\centering
\includegraphics[width=\linewidth]{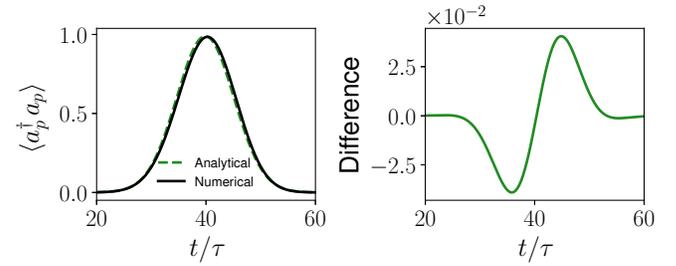}
\caption[A comparison between the number of photons in analytical and numerical results for $k_0=\pi/2D.$]{(left) The number of generated photons in a cavity for both full numerical, including cosines, and analytical results. (right) The difference between the results. The plots are for $2\sigma_+D = \sigma_-D=0.28$, $k_0=\pi/2D$, $\beta=2.2$, and $p=40$. }
\label{fig:full_Nu_Vs_An_NUMBER}
\end{figure}

Before exploring the new cases that were not possible to study using the analytic formalism developed in our previous work \cite{PhysRevA.100.033839}, we first compare some of our results from Ref.~\cite{PhysRevA.100.033839} with the results obtained using the numerical method discussed here. To facilitate the comparison between our previous results with the numerical result, we introduce $\sigma_+\equiv  \mathscr{E}_+ $ and $\sigma_- \equiv \mathscr{E}_- /(D \sin{(|k_0|D)})$.
In Fig.~\ref{fig:full_Nu_Vs_An_NUMBER} we compare the average number of photons from the full numerical results, including the cosines, with the analytical results obtained by approximating the cosines in \eqref{biphoton_03} to first-order about $k_0=\pi/2D$. As can be seen, limiting the expansion of the cosine functions to first order, when evaluating \eqref{biphoton_03} for $k_0=\pi/2D$, only results in a very small shift in the result to later time. This shift is due to the small change in the group velocity. For $k=k_o +\delta k$, when $\delta k \ll \pi/D$, then the group velocity is approximately given by $v_g(k)=v_g(k_0)(1-(\delta k)^2/2)$, where $v_g(k_0)=\omega_P \beta_1 D$. We thus see that the shift in group velocity is always negative for $k$ close to $k_0=\pi/2D$. Note that this approximation is only valid in a small region about $k_0=\pi/2D$, where the dispersion relation is almost linear. The importance of evaluating SVD numerically becomes clear when we move away from $k_0=\pi/2D$. Choosing $k_0$, which can have any value between $0$ to $\pi/D$ and is determined by the central frequency of the pump, can cause a significant change in the results. It is therefore important to be able to study the cases in which $k_0$ is not necessarily located at the linear region of the dispersion relation, or more specifically, the cases where $k_0\neq\pi/2D$.\par

To reasonably estimate the pump parameters we consider the case where $\sigma_{+}D=\sigma_{-}D=0.47$. This is equivalent of having a pump with temporal and spatial FWHM of $295\,\text{fs}$ and $3.3\,\mu \text{m}$, respectively. We choose the CROW material to be $\text{Al}_{0.35}\text{Ga}_{0.65}\text{As}$ due to its high nonlinearity and relatively large bandgap. In addition, we choose the pump wavelength to be $\lambda_S=775\,\text{nm}$, which not only results in generating counterpropagating signal and idler photons at the telecommunication wavelength, $\lambda_F=1550\,\text{nm}$, but also ensures operation below the band gap of $\text{Al}_{0.35}\text{Ga}_{0.65}\text{As}$. Choosing the periodicity of the CROW structure to yield a signal central wavelength of $1550\,\text{nm}$, gives $D\approx0.9\,\mu \text{m}$. As shown in Ref.~\cite{PhysRevA.100.033839}, choosing these parameters gives $\bar{\chi}_2\approx\chi_2/n^2(\omega_F)$, where $\chi_2\approx 100~\text{pm/V}$, appropriate for AlGaAs alloys~\cite{Yang:07,Gili:16,Carletti:15}, and $n\approx3.4$ at $\omega_{F}$. We now seek to determine the approximate number of pump photons under the above conditions that will give a squeezing parameter of $2.2$. Employing Eqs.~(\ref{eq:Q_0}) and using $Q_0 = \sqrt{2/(\pi \sigma_- \sigma_+)}$, the average number of photons in the pump ($|\alpha|^2$) is found to be $7.4\times10^{10}$, which gives a total pump pulse energy of approximately $19~\text{nJ}$.  We note that all of the above pump characteristics are easily achievable from a Ti:Sapphire laser.    \par 

\begin{figure}[!h]
\centering
\includegraphics[width=\linewidth]{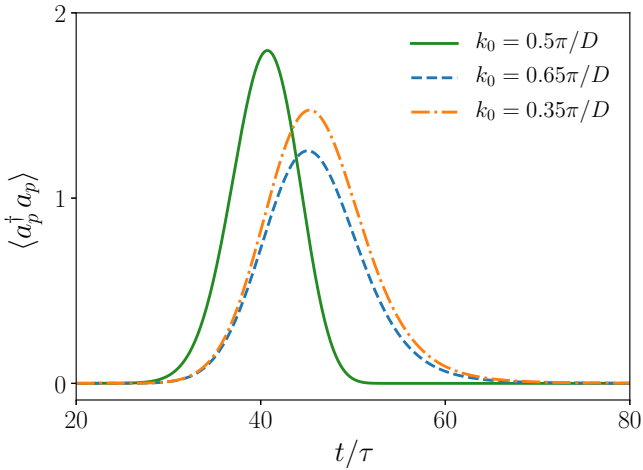}
\caption[Average number of photons in a cavity as a function of time for different $k_0$.]{Average number of photons in a cavity as a function of time for different $k_0$. The plots are for $\sigma_+D = \sigma_-D=0.28$, $\beta=2.2$, and $p=40$. }
\label{fig:different_k0_ph_number}
\end{figure}

In Fig.~\ref{fig:different_k0_ph_number}, we present the results for the average number of photons in a cavity as function of time for three different cases: $k_0=0.5\pi/D$, $0.65\pi/D$, and $0.35\pi/D$. In order to clearly demonstrate the effects of choosing $k_0$ on the results, we keep the other parameters to be identical in all the three cases. As can be seen, light propagates faster in the CROW when $k_0=0.5\pi/D$ and has the same velocity for the other two cases. This is exactly what one expects, since according to the dispersion relation, shown in Fig.~\ref{fig:dispersion}, the maximum group velocity occurs at $k_0=0.5\pi/D$ and it can be shown that the group velocities at the other two cases are equal. This effect, of course, could have been modelled using our analytic method, with just a different $k$. However, there is another obvious effect that could not. We see that for $k_0=0.65\pi/D$ and $0.35\pi/D$, the pulses are broadened.  This is the result of group velocity dispersion, which cannot be included using our previous analytic approach. Another important point extracted from Fig.~\ref{fig:different_k0_ph_number} is that although the group velocities are the same at $k_0=0.65\pi/D$ and $0.35\pi/D$, they yield different maximum number of photons. The reason behind this can be explained by Fig.~\ref{fig:dispersion}, where we plot the Q of the CROW structure as a function of the Block vector. As can be seen, the Q of the CROW depends on the Bloch vector, which in this case has a higher value at $k_0=0.35\pi/D$ than at $k_0=0.65\pi/D$. One might be tempted to go to an even smaller $k$ value, since the $Q$ is largest close to $k=0$.  However, as you approach $k=0$, the group velocity goes to zero, which means that the loss as a function of propagation distance will diverge.    
\par

\begin{figure}[!h]
\centering
\includegraphics[width=\linewidth]{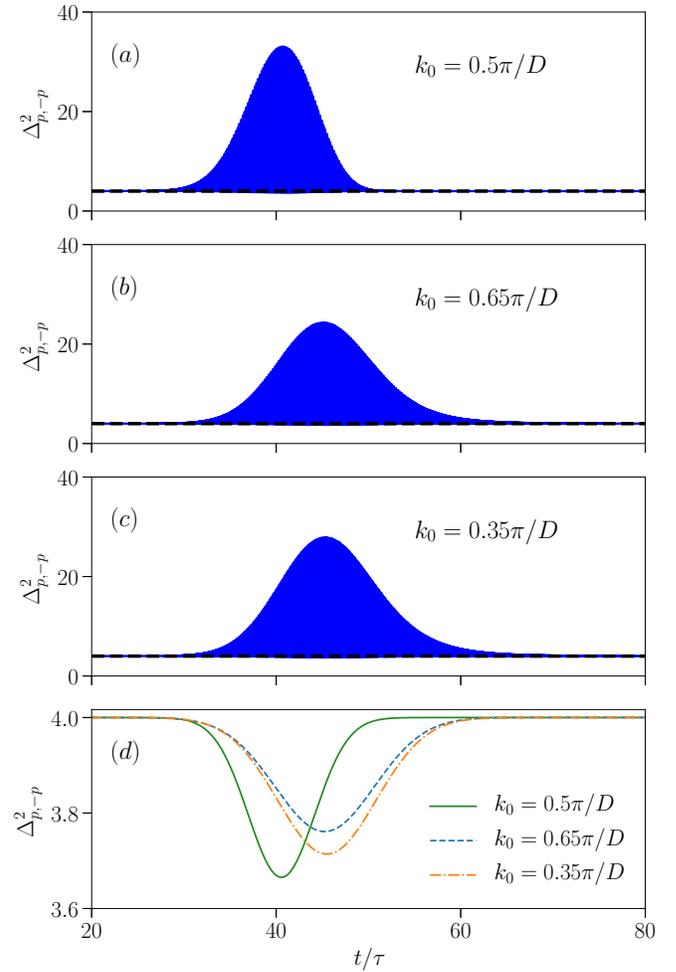}
\caption[Correlation variance between a pair of cavities in the CROW as a function of time for different $k_0$.]{Correlation variance between a pair of cavities in the CROW as a function of time for (a)~$k_0=0.5\pi/D$, (b)~$k_0=0.65\pi/D$, and (c)~$k_0=0.35\pi/D$. The lower envelopes are separately plotted in (d). The plots are for $\sigma_+D = \sigma_-D=0.28$, $\beta=2.2$, and $p=40$. }
\label{fig:different_k0_ENT}
\end{figure}

\begin{figure}[!hbt]
\includegraphics[width=\linewidth]{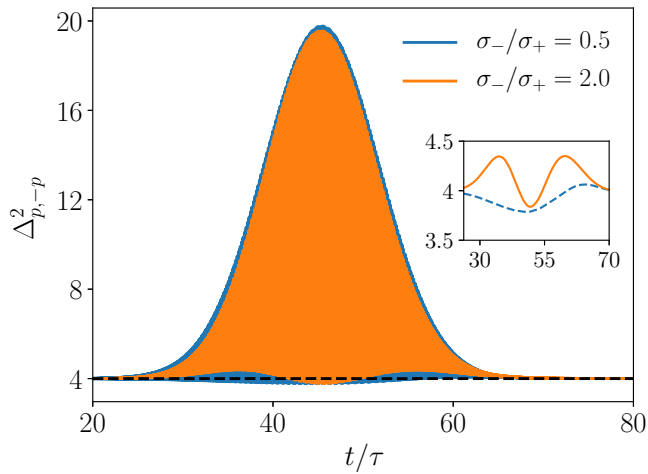}
\caption[Correlation variance between a pair of cavities in the CROW as a function of time for $k_0=0.35\pi/D$.]{Correlation variance between a pair of cavities in the CROW as a function of time for $k_0=0.35\pi/D$, $\beta=2.2$, and $p=40$. The inset shows the lower envelopes.}
\label{fig:Ent_different_ratio}
\end{figure}

In Fig.~\ref{fig:different_k0_ENT}, we compare the correlation variances for these three different cases with the same set of parameters considered in Fig.~\ref{fig:different_k0_ph_number}. From this figure, one sees that it also mirrors many of the same basic features such as the broadening and having slower velocity when $k_0\neq \pi/2D$. However, although these two figures demonstrate the effect of $k_0$ on the the average number of photons in a cavity and the correlation variance between the photons in a pair of cavities, it is hard to extract precise values for quantities such as the FWHM, the maximum number of photons, and the deviation of the minimum of the correlation variance from the inseparability threshold of $4$. Therefore, we summarize these quantities in Table~\ref{table:chapter5} for different $\sigma_-/\sigma_+$ ratios. As can be seen, having less intrinsic loss in the system when $k_0=0.35\pi/D$ comparing to $k_0=0.65\pi/D$ increases the deviation of the correlation variance significantly from inseparability value of 4. As can be seen, the highest average number of photons and correlation between the photons happens when $\sigma_-=\sigma_+$. Moreover, although for each $k_0$, according to Table~\ref{table:chapter5}, the system shows an almost identical behaviour for $\sigma_-=2\sigma_+$ and $2\sigma_-=\sigma_+$, there is an important difference. To demonstrate the difference between these two cases, in Fig.~\ref{fig:Ent_different_ratio} we plot the calculated correlation variances. As can be seen, when $\sigma_-/\sigma_+=0.5$ the inseparability criteria is met for a longer time than when $\sigma_-/\sigma_+=2.0$. Such a difference could not be revealed with analytical results due to the limitation in exploring the pumping configurations with $\sigma_+ > \sigma_-$ and shows that it is preferable to have $\sigma_-\leq \sigma_+$.  \par

\begin{table}
\caption{The maximum number of photons, FWHM of $\left\langle a_{p}^{\dagger}(t)a_{p}(t)\right\rangle$ (in units of $t/\tau$), and the deviation of the minimum of the correlation variance from the inseparability threshold of $4$ in a lossy system for $\beta=2.2$. The different pumping configurations A, B, and C represent the cases in which $\sigma_-D=\sigma_+D=0.28$, $\sigma_-D=2\sigma_+D=0.28$, and $2\sigma_-D=\sigma_+D=0.28$, respectively.}

\label{table:chapter5}
 \begin{tabular}{c  c  c  c  c} 
 \hline 
& $k_0$ & $\left\langle a_{p}^{\dagger}a_{p}\right\rangle_{max}$ &  $4-\left(\Delta_{p,-p}^{2}\right)_{min}$ & FWHM \\  [1ex]
 \hline
A &  & 1.80 & 0.35 & 8.58\\ 
\textcolor{black}{B} & \textcolor{black}{$0.50\pi/D$} & \textcolor{black}{0.99} & \textcolor{black}{0.30} & \textcolor{black}{12.39}\\ 
\textcolor{black}{C} &  & \textcolor{black}{0.99} & \textcolor{black}{0.31} & \textcolor{black}{12.33}\\ 
 \hline
A &  & 1.26 & 0.25 & 11.98\\ 
\textcolor{black}{B} & \textcolor{black}{$0.65\pi/D$} & \textcolor{black}{0.81} & \textcolor{black}{0.19} & \textcolor{black}{14.99}\\ 
\textcolor{black}{C} &  & \textcolor{black}{0.82} & \textcolor{black}{0.20} & \textcolor{black}{15.02}\\ 
 \hline
A &  & 1.48 & 0.29 & 12.15\\ 
\textcolor{black}{B} & \textcolor{black}{$0.35\pi/D$} & \textcolor{black}{0.96} & \textcolor{black}{0.23} & \textcolor{black}{15.07}\\ 
\textcolor{black}{C} &  & \textcolor{black}{0.97} & \textcolor{black}{0.23} & \textcolor{black}{15.08}\\ 
 \hline 
 \end{tabular}
\end{table}

\par

\section{Conclusion}
\label{sec:comclusion}
In this work, we applied the numerical Schmidt decomposition method to the general formalism developed in our previous work~\cite{PhysRevA.100.033839}. This allowed us to investigate the validity of the approximations made in deriving analytical time dependent expressions for the average number of photons in each cavity and the correlations between cavities presented in Ref.~\cite{PhysRevA.100.033839}. Moreover, using numerical methods, we were able to investigate more general cases and overcome the limitations we had set to derive the analytical expressions. Furthermore, the numerical results presented in this work opened up the possibility of studying more general pumping configurations as well as the structures in which the dispersion relation is not simply given by the nearest-neighbor tight-binding approximation. for instance, using FDTD with periodic boundary conditions, it has been shown that it is possible to determine the Bloch functions and the exact dispersion relation for different waveguides. Such a method can be applied to CROWs that have more complicated dispersion than given by Eq.~\ref{eq:dispersion_NNTB}. In addition, it is possible to employ the Wannier functions to investigate the properties of different PhC structures such as PhC cavities and PhC defect waveguides without going through a tight-binding parametrization. Employing these two features of Wannier functions alongside the numerical results presented in this work provide us a better tool to study more complicated structures. 

\section{Funding Information}

This work was supported by Queen’s University and the Natural Sciences and Engineering Research Council of Canada (NSERC).


\bibliography{sample}

\begin{thebibliography}{10}
\bibitem{complex}
Throughout this work, we place a tilde on top of quantities that are inherently
  complex (as opposed to real).
\end{thebibliography}

\bibliographyfullrefs{sample}


\end{document}